# Surface declination governed asymmetric sessile droplet evaporation

**Purbarun Dhar** [a, #]**, Raghavendra Kumar Dwivedi** [b]**, and A R Harikrishnan** [c, *]

[a] Department of Mechanical Engineering, Indian Institute of Technology Kharagpur, Kharagpur–721302, West Bengal, India

[b] Department of Mechanical Engineering, Indian Institute of Technology Ropar, Rupnagar–140001, Punjab, India

[c] Department of Mechanical Engineering, Birla Institute of Technology and Science Pilani, Pilani–333031, Rajasthan, India

[#] *Corresponding author:*

[#] E–mail: purbarun@mech.iitkgp.ac.in ; purbarun.iit@gmail.com

[#]Phone: +91-3222-28-2938

*E–mail: ar.harikrishnan@pilani.bits-pilani.ac.in

## Abstract

The article reports droplet evaporation kinetics on inclined substrates. Comprehensive experimental and theoretical analyses of the droplet evaporation behaviour for different substrate declination, wettability and temperatures have been presented. Sessile droplets with substrate declination exhibit distorted shape and evaporate at different rates compared to droplets on the same horizontal substrate and is characterized by more often changes in regimes of evaporation. The slip-stick and jump-stick modes are prominent during evaporation. For droplets on inclined substrates, the evaporative flux is also asymmetric and governed by the initial contact angle dissimilarity. Due to smaller contact angle at the rear contact line, it is the zone of a higher evaporative flux. Particle image velocimetry shows the increased internal circulation velocity within the inclined droplets. Asymmetry in the evaporative flux leads to higher temperature gradients, which ultimately enhances the thermal Marangoni circulation near the rear of the droplet where the evaporative flux is highest. A model is adopted to predict the thermal Marangoni advection velocity, and good match is obtained. The declination angle and imposed thermal conditions interplay and lead to morphed evaporation kinetics than droplets on horizontal heated surfaces. Even weak movements of the TL alter the evaporation dynamics significantly, by changing the shape of the droplet from ideally elliptical to almost spherical cap, which ultimately reduces the evaporative flux. The life time of the droplet is modelled by modifying available models for non-heated substrate, to account for the shape asymmetry. The present findings may find strong implications towards microscale thermo-hydrodynamics.

*Keywords:* evaporation; sessile droplet; inclined surface; wettability; stick-slip phenomenon; heat transfer; PIV## 1. Introduction

Droplet evaporation, a diffusion driven mechanism due to the vapour concentration gradient existing between the surface of the droplet and the ambient far-field. This is significant for a wide range of applications, such as automobile industries where engine performance is governed by droplet atomization and its combustion characteristics [1, 2], spray cooling processes [3-5], ink-jet printing [6], etc. It is also of importance in medical therapeutics and diagnostics like tissue occlusion in cancer treatment [7], and patterning of biological fluids as a simple and prompt diagnosis to identify diseased conditions [8]. Droplets are categorized as pendent and sessile droplets, where the latter rests in equilibrium on a solid surface, while in equilibrium with the gas phase surrounding it. The nature of interfacial interactions between the three phases determines the contact radius and the contact angle of the droplet. A pioneering and fundamental study on the evaporation of sessile droplets by Picknett and Bexon [9] showed that the droplet will evaporate via two major mechanisms; the constant contact radius mode (CCR) and the constant contact angle mode (CCA). In the CCR mode, the wetting diameter remains constant and contact angle diminishes (pinning mode), whereas in the CCA mode, the contact angle remains constant and the wetting diameter retreats from the initial value (depinning mode).

Literature reports also discuss a third mode of evaporation, the mixed mode or stick-slip mode, where the droplet evaporates with diminishing wetting diameter and variations in contact angle [10]. Bormashenko et al. [11] showed that low surface energy substrates exhibit weak pinning behaviour and evaporates by the stick-slip mode; whereas high surface energy substrates exhibit strong pinning and evaporate mostly by the CCR mode. The work also modelled the motion of the triple line (TL) in terms of the substrate energy as developed by Shanahan [12]. Transients of the droplet contact diameter and the contact angle will ultimately depend on the evaporative flux, and vice versa. Recent studies mostly focus on motion of the TL [13-18] and the general conclusion is that the substrate tribology and wetting properties are the dominating factors for the TL dynamics. Another work [19] has reported a fourth mode of evaporation, the stick-jump mode. The work experimentally showed the occurrence of sudden jumps in the wetting diameter and the contact angle when a droplet is deposited on a smooth substrate and begins to evaporate. A similar phenomenon is also reported by numerical simulations [20] of droplet evaporating on a chemically patterned surface.

Deegan et al. [21] observed highest value of evaporative mass flux in the neighbourhood of the TL region for a sessile droplet on horizontal substrate. When a droplet rests on a hydrophilic inclined substrate, it is under the effect of gravity acting along the droplet. The force along the inclined plane in the downward direction results in distortions of the shape of the droplet. Due to this asymmetric shape, the equilibrium contact angle can be resolved as the front ($\theta_f$) and rear ($\theta_r$) contact angles. The asymmetric drop shape thereby induces asymmetry in the evaporative flux. Reports [22] have experimentally and numerically evaluated the deposition pattern on inclined surfaces where radially non-uniform deposition was noted. The deposition strength was found to be a direct function of the droplet initial volume and declination angle. Extrand and Kumagai [23] concluded that the chemical nature of the substrates influences the contact angles on inclined planes rather than surface roughness, and the retention forces increases the elongation of the droplet profile when deposited. Lattice Boltzmann simulations [24] of droplets on inclined smooth and topographic substrates of different wettability show the transients of partial pinning and displacement of the TL before the critical retention force is achieved. Investigations [25] have reported the shape of a sessile drop resting on an inclined substrate based on the energy minimization principle. Experiments [26] show that the droplet life time decreases with increase in declination angle (0° to 90°).

In the present study, we probe the evaporation mechanism of sessile water droplets on three substrates (glass, aluminium and Teflon) of different wettabilities, for different surface declinations (0°, 30°, 60° and 90°). The evaporation kinetics is experimentally studied, and the evolution of droplet volume, droplet radius and contact angle is deduced. The non-uniform evaporative flux along the non-axisymmetric droplet at various declination angles is theoretically predicted and correlates with the experimentally observed evaporation kinetics. Flow visualisation within the droplets has been conducted to further probe the mechanisms behind the observed evaporation kinetics. The substrate declination is found to drastically influence the contact line dynamics and pinning–depinning phenomenon during evaporation, which in turn has a direct correlation with the life time of the droplet. A theoretical analysis has been presented to model the underlying physics, by coupling the evaporation rate with the contact line dynamics for surfaces of various declination and wettability. Additionally, the effect of substrate temperature is probed and its role towards modulating the contact line dynamics in asymmetric droplets is revealed. The findings may shed important insights on evaporation kinetics of droplets with distorted and asymmetric shapes.

## 2. Experimental methodology

The schematic of the experimental setup has been illustrated in figure 1. The setup consists of a precision droplet dispensing mechanism controlled by a digitized controller. A monochromatic CCD camera (Holmarc Opto-mechatronics, India), recording at 1280 x 960 pixels resolution, and at 10 fps, attached with a long distance microscopic lens is used for image acquisition. A light emitting diode (LED) array (Holmarc Opto-mechatronics, India) is used as the illumination source. The droplet dispenser (accuracy of ~ 0.1 µl) is loaded with a 50 µl glass chromatography syringe, having a stainless steel needle (22 gauge), which dispenses the droplet on the substrate from very close proximity. Droplet volume of ~20 µl is used in all the present experiments. This leads to sessile droplets with Bond number (ratio of gravitational to surface tension force) of ~ 1-2, which are equally under the influence of gravitational and surface forces. This ensures the formation of asymmetric tear-shaped droplets due to surface declination. The substrate in question is mounted on a heating unit, whose inclination can be varied accurately from 0–180$^o$ via a digital actuator. The temperature of the heater is maintained with the help of a PID controller unit. The temperature of the substrate is measured by a thermocouple, which also acts as the feedback input to the heater controller for automatic cut-off and start of heating.

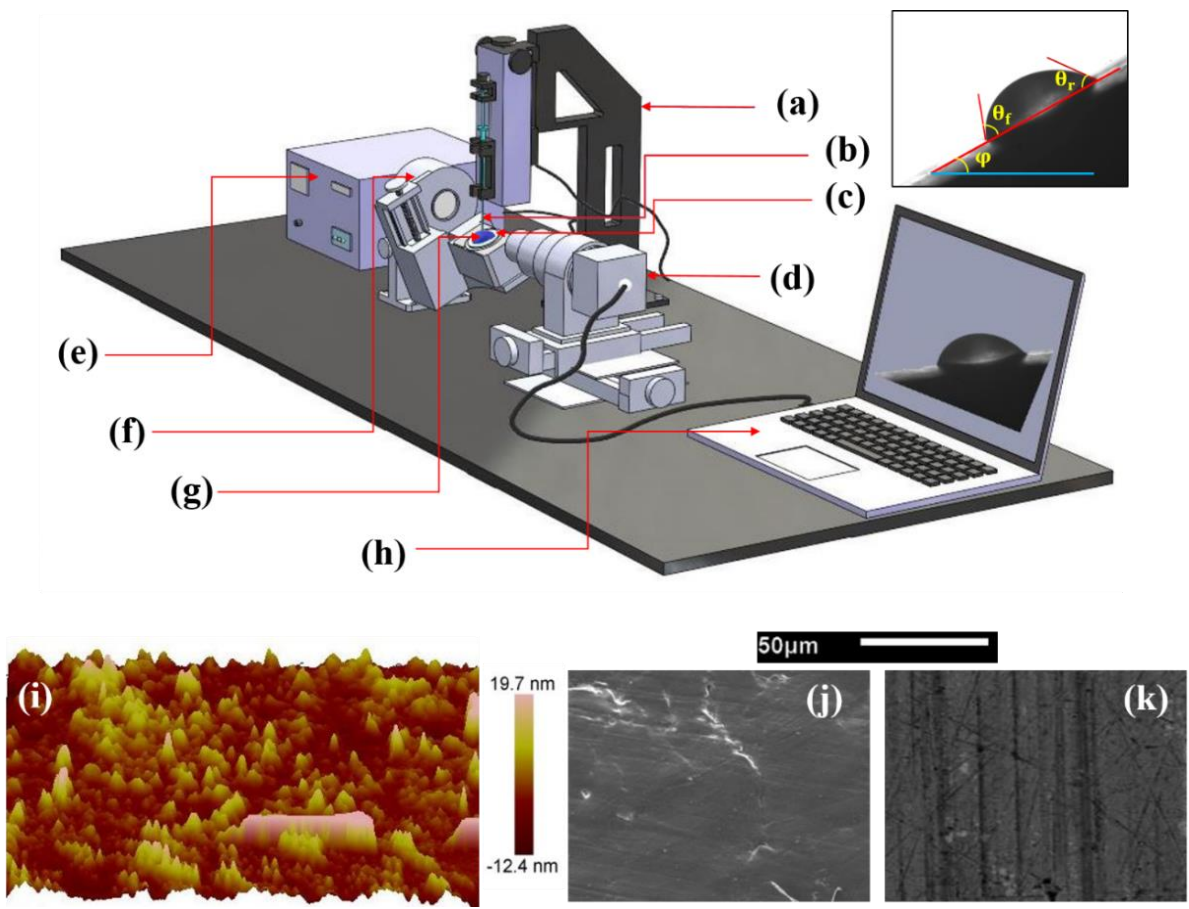

**Figure 1:** Diagram of experimental setup (a) droplet dispenser (b) syringe (c) substrate fit on heater with inclination mechanism (d) CCD camera with long distance microscopic lens and 2-axis movement (e) substrate inclination and heater controller (f) LED illumination source (g) droplet (h) data acquisition computer, (i) AFM image of the glass slide (j) SEM image of Teflon and (k) aluminium substrate. The inset shows the droplet on inclined substrate with front and rear contact angles.

The whole setup is housed within an acrylic chamber and placed on a vibration free horizontal table to suppress all ambient disturbances. The chamber is equipped with a digitized thermometer and hygrometer to note the temperature and humidity conditions within the chamber, at 20 mm away from the droplet. For all the set of experiments, the temperature was observed to vary as 31 ± 2 ºC, and the relative humidity as 49 ± 4%. The wettability effect is studied using three substrates, viz. glass, aluminium and Teflon. The surface roughness contours of the substrates are determined using atomic force microscopy (AFM) and scanning electron microscopy (SEM) (refer fig. 1). The average surface roughness of the substrates are determined using a surface profilometer and are tabulated in table S1 (refer supporting information). All the experiments are performed for substrate declination angle of φ=0º, 30º, 60º and 90º from the horizontal plane. Each set of experiment is repeated thrice. The recorded images are post processed in the open source software ImageJ to obtain the transient variation of volume, contact diameter, droplet height, and contact angle.

Particle image velocimetry (PIV) (not shown in the experimental setup) is conducted to quantify the dynamics of circulation, if any, inside the droplet during the evaporation and to understand its effect on evaporation kinetics. A continuous wave laser (532 nm, 5 mW, Roithner GmbH, Germany) is used as the illuminating source and a cylindrical lens has been used to generate a light sheet of thickness ~0.5 mm. The light sheet is focussed vertically and illuminates the vertical mid-plane of the sessile droplet. Fluorescent, neutrally buoyant (with water at 300 K), inert spheres of polystyrene of 10 μm diameter (Cospheric LLC, USA) are utilized as seeding particles. The PIV images are captured at 20 fps using the CCD camera, with typical resolutions of ~120 pixels/mm. A four pass cross correlation algorithm has been employed in post processing the images using the open source code PIVLab. Interrogation window sizes of 64, 32, 16 and 8 pixels in consecutive passes were employed to obtain better signal to noise ratios. PIV is only done for droplets on Teflon substrates as the surface is not largely wetting as glass, and good imaging is possible. Due to difficulty in positioning the laser close to the diffusing Teflon surface, the PIV studies are limited to declination angles of 0°, 30°, 45°, and 60°.

## 3. Results and discussions

### 3. a. Evaporation kinetics at different substrate declinations

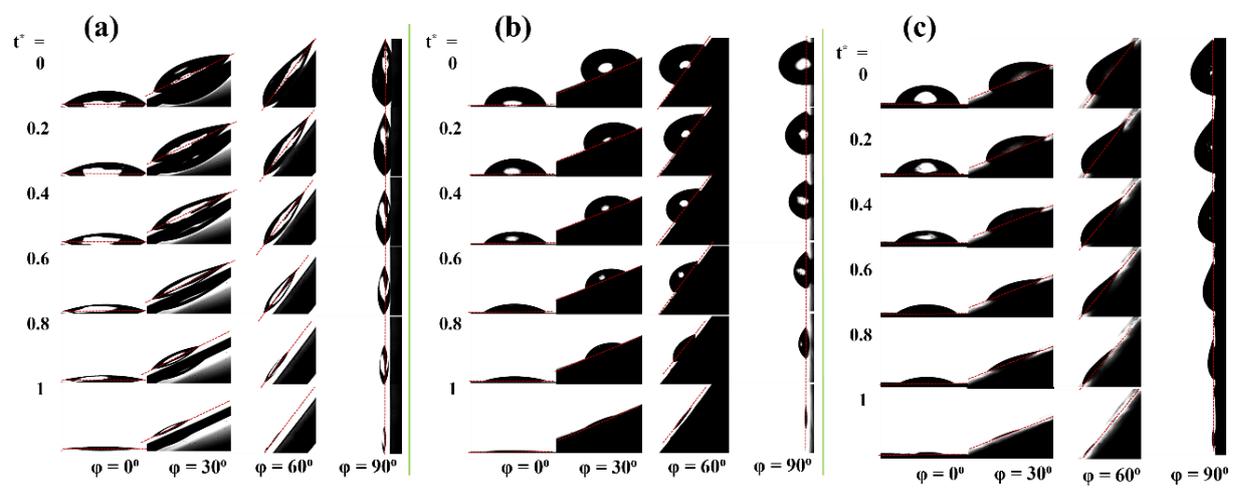

**Figure 2:** Snapshots of time transients of an evaporating sessile droplet on (a) glass, (b) aluminium and (c) Teflon substrates at different declination angles, where $t^*=t/t_f$ is the non-dimensional time (where t is the evaporation time elapsed and $t_f$ is the droplet life time). The droplet on glass, at zero declination and $t^*=0$ acts as the scale bar for the whole figure. The contact diameter of the droplet is 2.8 mm. The thin red lines are guide for the eyes to demarcate the original droplet from the substrate surface.

The transients of the evaporation kinetics on glass, aluminium and Teflon, at different declinations, and at different non-dimensionalized droplet life times have been illustrated in fig. 2. It is observed that the surface wettability and declination together lead to the asymmetric shape, which in turn modulates the droplet shape evolution as evaporation progresses. The force balance (gravitational and surface tension force) for a sessile droplet resting on an inclined surface is expressed as

$$mg\sin(\varphi) = k\gamma_{lv}\left(\cos(\theta_r) - \cos(\theta_f)\right)d \qquad (1)$$

Where, m is the mass of the droplet, $\varphi$ is the declination angle, d is the instantaneous wetting diameter, $\theta_r$ and $\theta_f$ are the instantaneous rear and front contact angles, respectively, $\gamma_{lv}$ is the surface tension of the liquid-vapour interface and *k* is a non-dimensional constant which accounts for the elongation of the droplet [27]. For the same volume of droplet, if the inclination angle is increased, a situation arrives where onset of sliding sets in, at a critical inclination angle ($\varphi_{cri}$). At this instant, the front and rear contact angles manifest as the advancing ($\theta_{adv}$) and receding ($\theta_{rec}$) contact angles, respectively. Under this situation, the RHS of equation (1) is called the critical retention force or maximum retention force and is expressed

$$F_{cr} = k\gamma_{lv}\left(\cos\theta_{rec} - \cos\theta_{adv}\right)d \qquad (2)$$

If the gravitational force exceeds the maximum retention force, the droplet will slide. However, in the present study, a partial depinning [28-30] of the droplet occurs in majority of the cases, where the rear edge of the TL is sliding and the front edge remains pinned throughout the evaporation process.

Figure 3 illustrates the variation of droplet volume with progressing evaporation, for droplets on glass surfaces of variant declinations. In the present approach, a modified version of the spherical cap geometry was used to determine the droplet volume. In reality, the droplet is asymmetric and does not conform to the spherical cap shape. In the present case, the spherical cap volume is evaluated twice using the droplet geometrical parameters, once using $\theta_f$ and another using $\theta_r$. The effective volume is determined as the average of the two values. It is noted that this approach leads to estimation of the initial droplet volume on all substrates and declinations within ± 8% of the experimentally dispensed volume. The non-dimensional volume raised to the power of 2/3 is plotted [31] illustrate the role of declination on the evaporation kinetics. It is noted that on a horizontal surface, the trend of has a nearly linear nature. With increase in the declination angle to 30°, a certain degree of non-linearity is induced, which signifies that the volume flux from the droplet is no longer constant with time, and this is caused by the asymmetric droplet shape. The non-linearity is furthered with increase in the declination angle, which is caused by the augmented asymmetry in the droplet shape. A qualitative idea of this behaviour and droplet shape asymmetry can be noted from fig. 2.

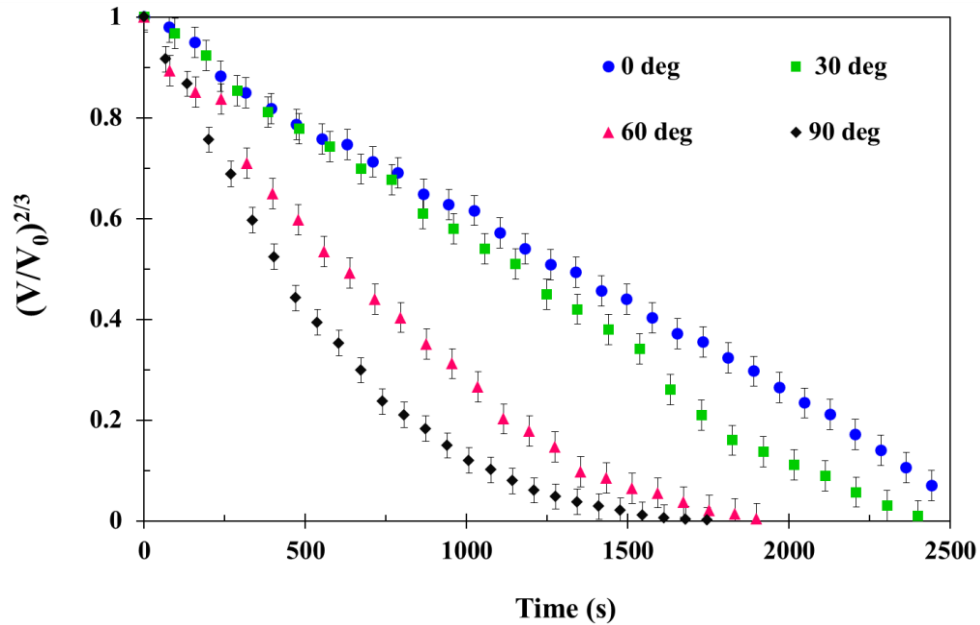

**Figure 3:** Transient evolution of the non-dimensional droplet volume for sessile droplets on glass substrates of different declinations.

Figure 4 illustrates the dynamics of the contact diameter and contact angles of a droplet evaporating on a glass substrate at different declination angles. On horizontal glass substrate the droplet evaporates via three modes: CCR, CCA and mixed mode. With increase in the declination angle, the droplet de-pins earlier. This behaviour is governed by the combined effect of the dissimilarity in contact angles and the gravitational force. With increase in the declination, the force component along the inclined plane increases, which enhances the front contact angle and decreases the rear contact angle; resulting in a tear shaped droplet. For droplets evaporating on horizontal glass surface, the TL receeds continuously from both the ends after the CCR mode. This is follwed by the mixed regime, and finally the prcess ends with the CCA regime. With declination, the mixed and the CCA modes compete with each other and appear multiple times, one after the other. At $30^o$ declination, a jump-slip mode is observed, which is due to the interplay of the retention force and the gravitational force adjusting themselves during the shrinkage of the droplet volume with time. This interplay leads to transient pinning-depinning phenomena, leading to the CCA and mixed modes occuring repetatively during the evaporation of the tear shaped droplets.

The TL dynamics during various declination angles are closely correlated with the initial contact angle dissimilarity at that particular declination. This behaviour is evident from the analysis of the values presented in table S2 (supporting information) along with the TL dynamics illustrated in figure S1 (refer supporting information). Figures S2 and S3 (supporting information) illustrate the contact diameter and contact line dynamics during the evaporation of droplets on Teflon and aluminium substrates of different declinations, which gives further insight on the dynamics of evaporation and the TL on substrates with declination, and the role of wettability. The role played by wettability of the surface is manifested by the contact angle hysteresis of a particular surface-droplet pair. It is noted that if the value of $\theta_f$ lies between the advancing and receding angles, then the front edge of the droplet will remain pinned during evaporation (as mostly observed in the present study). Simultaneously, if $\theta_r$ reduces below the receding angle, then the rear edge of the droplet will move towards the front edge, leading to slip behaviour. Additionally, if $\theta_f$ increases above the advancing contact angle, then the front edge of the droplet moves along the inclined place, causing elongation in the wetting diameter. Typical slip-jump behaviour is observed on the glass surface at $\varphi=30°$ and the dynamics has been illustrated in fig. S1.

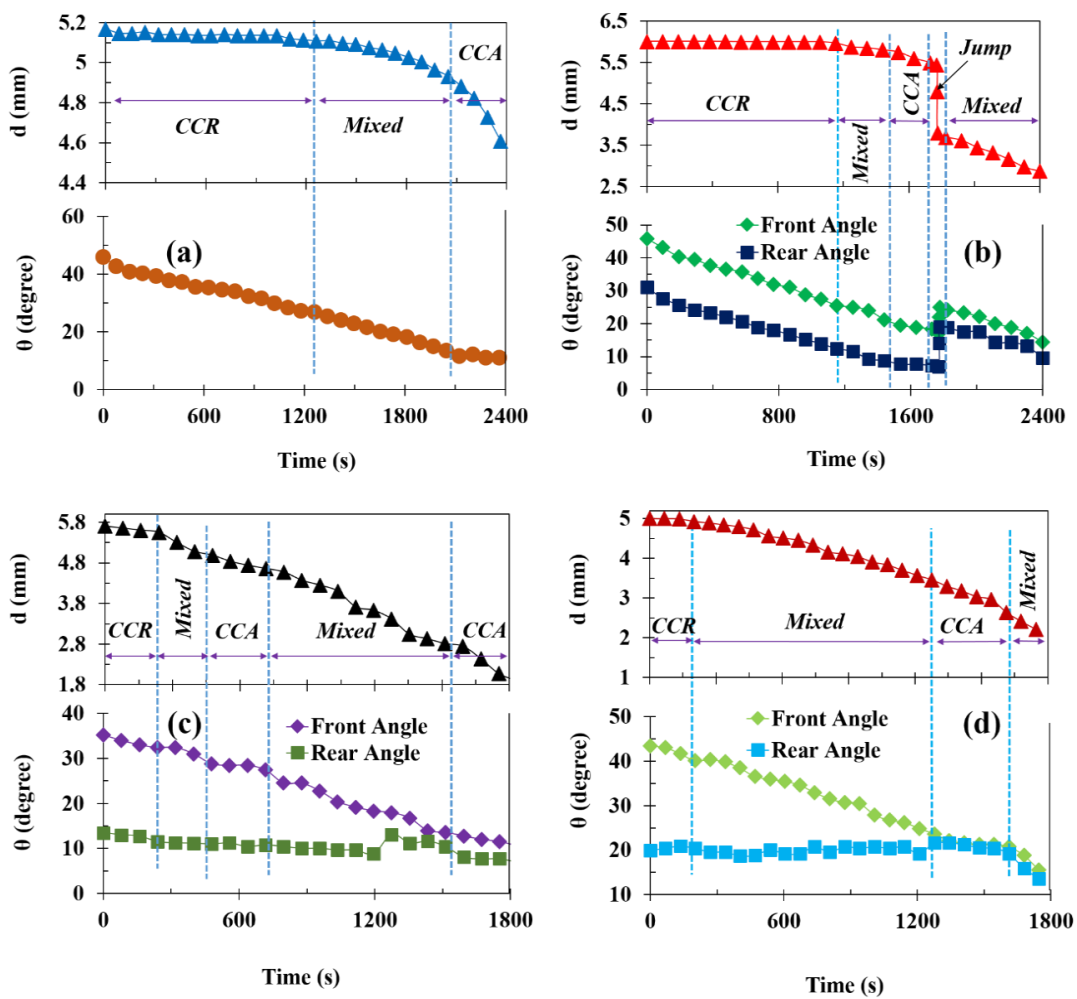

**Figure 4:** Evolution of contact diameter and contact angle on glass substrate at $\varphi=$ (a) 0⁰ (b) 30⁰ (c) 60⁰ and (d) 90⁰. The typical regimes noted have been labelled in the figure.

For a particular declination angle, initially the droplet remains pinned and after certain time period it recedes. The asymmetric evaporative flux from the tear shaped droplets induces dissimilar rates of change of $\theta_f$ and $\theta_r$. For zero declination, the rate of change of the two contact angles is equal, and hence both the ends of the TL show same dynamics (on a homogeneous and uniform surface). With increasing declination on the non-metallic substrates (glass and Teflon), the trend of reduction of $\theta_f$ remains similar to that of the horizontal surface. However, the change of $\theta_r$ shifts from decreasing trend to almost constant trend as the declination increases. This behaviour thus results in decrease in the retention force of the droplet as evaporation progresses. This leads to increase in the propensity of pinning-depinning behaviour, which leads to the frequent CCA-mixed mode transitions noted on inclined surfaces (figs. 4, S2 and S3). In majority of the present experiments, partial depinning dominates over sliding of the droplet, as the front edge of the TL droplet remains pinned during the droplet lifetime. However, in case of glass and Teflon, a combination of partial depinning and sliding of the droplet at 60 and 90° declination is noted. Such behavior is however, absent in case of aluminium, and the front edge of the TL always remains pinned.

As the surface roughness of Teflon is higher than that of the glass substrate (Table S1, SEM and AFM images in fig. 1), the droplet exhibits greater propensity of pinning on the former. As reported by articles [32-34], higher surface roughness leads to greater frictional force at the droplet-substrate interface. Increase in surface roughness induces enhanced contact angle dissimilarity in tear-shaped droplets and ultimately improves the pinning force. Hence the onset of depinning occurs at a much later stage of evaporation compared to that on glass (discussed in subsequent sections). The contact angle dissimilarity on Teflon being higher than that on glass induces more non-uniform evaporative flux, which ultimately reflects on the contact line pinning-depinning mechanism. On Teflon the evaporation is more inclined towards the CCR and mixed modes, as the high roughness ensures the CCR due to high pinning propensity. On aluminium, the TL exhibits jump state at every declination angle studied. Further, both the contact angles decrease at a uniform rate unlike on glass and Teflon substrates. This could be attributed to the wetting behaviour of polished metallic surfaces, where such slip-stick phenomenon is triggered by the surface texture [35]. The role of wettability on the declination induced slip-stick behaviour has been illustrated in fig. S4.

### 3. b. Evaporative flux of asymmetric droplets

It has been noted from fig. 3 that the evaporation rates are enhanced due to declination. The next approach will be to understand how the evaporative flux is modulated by the surface declination, which leads to overall increment in the volume reduction rates. The analytical solution for the net evaporative flux in terms of contact angle ($\theta$) for a spherical cap approximated droplet is proposed [36] in Eqn. 1 (by treating the evaporation process to be quasi-steady). Using a Legendre function of first kind and implementing reported methodology [37], the evaporative flux $J(\alpha)$ is expressed as

$$J(\alpha) = J_0 \left( \frac{\sin(\theta)}{2} + \sqrt{2}(\cosh(\alpha) + \cos(\theta))^{3/2} \right) \left( \int_0^\infty \left( P_{i\tau-\frac{1}{2}} \frac{\tau\cosh(\theta\tau)}{\cosh(\pi\tau)} \tanh(\tau(\pi-\theta))\cosh(\alpha) \right) d\tau \right) \quad (3)$$

Where $J_0 = D(C_s - C_\infty)/R$ is the evaporative flux at $\theta=90°$ i.e. for a purely hemispherical sessile droplet. D is the vapour diffusion coefficient, $C_s$ and $C_\infty$ are the water vapour concentration at the interface of the droplet and far away from the droplet, respectively, R is the droplet radius, and $\theta$ is the contact angle. The term $P_{i\tau-1/2}(\cosh\alpha)$ is the complex Legendre function with hyperbolic argument. An approximate form of the evaporative flux [37] can be expressed from eqn. 3 as

$$J(\theta) = J_0 \left(0.27\theta^2 + 1.3\right) \left[ 0.6381 - 0.2239 \left(\theta - \frac{\pi}{4}\right)^2 \right] \left[1-\varsigma^2\right]^{\left(\frac{\theta}{4} - \frac{1}{2}\right)} \quad (4)$$

Where $\varsigma = r/R$ represents the normalized droplet radius.

The eqn. (4) is valid for $\varphi=0°$ (i.e. a typical spherical cap sessile droplet). As the surface declination influences or distorts the droplet symmetry, two contact angles, $\theta_f$ and $\theta_r$ appear (fig. 1 inset) and the direct application of eqn. (4) is not possible. Hence in the present study, we have used a new methodology to incorporate the effect of declination angle in the analytical expression for the evaporative flux. The asymmetric droplet profile has been split into two different symmetric, spherical cap segments. One such droplet his constructed based on $\theta_f$ and the other based on $\theta_r$. The construction of the two imaginary spherical cap droplets is constrained such that the total volume of the two droplets equates to the volume of the original droplet. The total evaporative flux is evaluated as the piecewise average of the evaporative flux from each segment. The resultant expression for the ratio of net effective flux is as

$$\chi(\varsigma) = \frac{J(\theta)}{J(\theta_{\varphi=0})} = \frac{\left(0.27\theta_\varphi^2 + 1.3\right) \times \left[0.6381 - 0.2239\left(\theta_\varphi - \frac{\pi}{4}\right)^2\right]\left[1-\varsigma^2\right]^{\left(\frac{\theta_\varphi}{4} - \frac{1}{2}\right)}}{\left(0.27\theta_{\varphi=0}^2 + 1.3\right) \times \left[0.6381 - 0.2239\left(\theta_{\varphi=0} - \frac{\pi}{4}\right)^2\right]\left[1-\varsigma^2\right]^{\left(\frac{\theta_{\varphi=0}}{4} - \frac{1}{2}\right)}} \quad (5)$$

Eqn. (5) correlates the evaporative flux at different declination angles $\varphi$, to the evaporative flux at $\varphi=0$ (i.e. typical sessile droplet evaporation, given by $J(\theta)$). $\theta_\varphi$ represents the effective contact angle at declination angle $\varphi$, and $\theta_{\varphi=0}$ is the contact angle at $\varphi=0$. The equation is solved numerically in Mathematica using the contact angle and vapour concentration conditions. The evaporative flux deduced from the adopted method is integrated over the droplet surface area and the volume loss is noted. These values are in agreement (within ~ ±15%) of the observations in fig. 3.

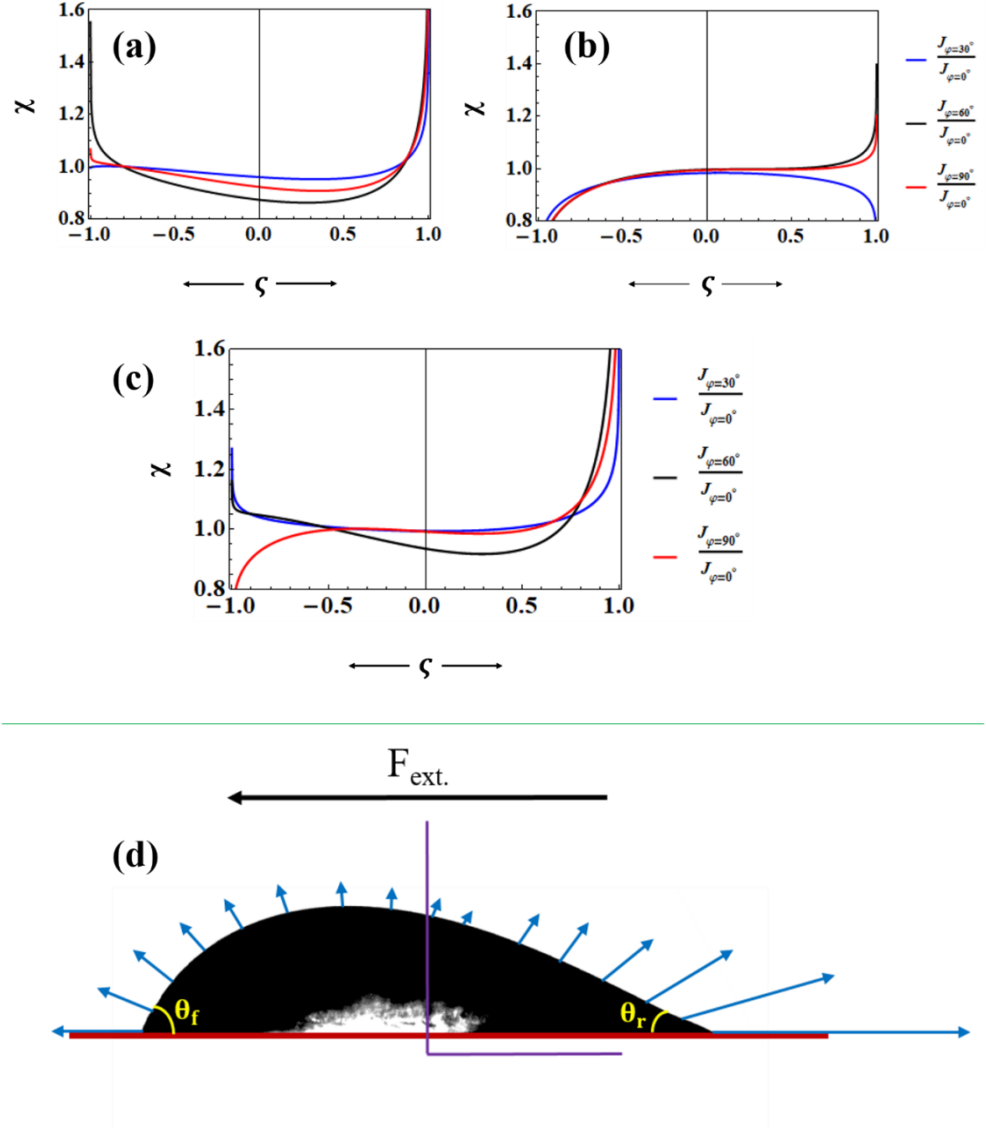

**Figure 5:** Variation of the ratio of evaporative flux ($\chi$ represented as $J_{\varphi=(30° \text{ or } 60° \text{ or } 90°)}/J_{\varphi=0}$) at different $\varphi$ with respect to normalised radius on (a) glass (b) aluminium and (c) Teflon. (d) Schematic illustration of the nature of variation of the magnitude of evaporative flux on Teflon substrate at $\varphi=60°$. The horizontal large arrow represents the direction of declination (labelled $F_{ext}$) represents the external force (gravitational) on the droplet. The blue arrows indicate the magnitude of the evaporative flux at a given point. In (a), (b) and (c) the blue, black and red lines represent $\chi$ at $\varphi = 30, 60$ and $90°$, respectively. $\varsigma=1$ represents the rear end of the axisymmetric droplet.

For symmetric droplet evaporation ($\varphi=0$) with acute contact angle, the evaporation rate is maximum at the TL, whereas for obtuse contact angles, it is maximum at the apex of the droplet. Fig. 5 illustrates the nature of variation of $\chi$ at different $\varphi$ on different substrates. The flux values correspond to the state of the droplets within the first few minutes of evaporation. A schematic case is illustrated in Fig. 2 (d) for the initial few minutes of a droplet evaporating on Teflon at $\varphi=60°$ ($\theta_f \sim 72°$ and $\theta_r \sim 23°$, fig. S2, supporting information). The arrows emanating from the droplet surface indicate the typical variation of χ and the arrow lengths indicate the strength of the evaporative flux. At the rear TL, at $\varsigma \sim 0.9999$ χ ~18.86, and at the front TL, at $\varsigma \sim -0.9999$, χ ~2.28. The minimum value of χ ~0.9375 appears at $\varsigma \sim 0.2257$. This indicates that both the front and rear contact points of the TL have higher evaporation rates than the analogous regions of φ=0 droplet. However, the tear shaped droplet leads to largely enhanced evaporation rate at the rear contact line, where the droplet exhibits largely reduced contact angle due to gravity induced draining of the liquid towards the direction of declination. However, in the region in the neighbourhood of $\varsigma \sim 0.2257$, the vaporization rate is lower than the φ=0 droplet. Additionally, the minima of the flux are shifted from the apex to the droplet due to the asymmetry of the shape. These minima can be located from fig. 5

On Teflon and aluminium, the droplet wets the surface less compared to glass, leading to larger droplet heights and smaller contact diameter than on glass (refer fig. 2). Consequently, with substrate declination, the front edge of the droplets on aluminium and Teflon bulge outwards further than on glass, especially for the 90º case of Teflon and all declinations of aluminum. Consequently, the Stefan flow around the evaporating droplet [31, 38] near the front edge of the TL is disrupted due to the physical bulge of the distorted droplet shape. This in turn leads to reduction in the evaporative flux at the front edge compared to the rear edge, and this behavior is noted from fig. 5 for the said cases. On inclined droplets, the front edge zone of the TL where the evaporation rate is less, remains pinned always. Whereas at the rear edge zone of high evaporative flux, the TL may stick, slide or jump depending on the substrate wettability and the declination angle. On glass at φ=30⁰, a sudden jump in the wetting diameter and contact angles appears consistently. Sudden jump transitions imply large imbalance of capillary and gravitational force. A large difference in the values of the evaporative flux between the rear and front edges of the TL induces the jump transition to thermodynamically stabilize the droplets. During the jump transition, the increase in $\theta_r$ (~6° to ~22°) is larger than the increase in $\theta_f$ (~18° to ~24°), which justifies the above reasoning. The difference of the evaporative flux for aluminium is smaller than glass, and consequently it exhibits jump transitions which are not so rapid, but slower.

**3. c. Shape asymmetry mediated internal thermo-hydrodynamics**

Evaporation of sessile droplets is known to generate internal advection due to thermal Marangoni effect caused by the non-uniform evaporative cooling of the droplet. Evaporating sessile droplet on horizontal glass substrate exhibits generation of internal Marangoni circulation cells, and the directionality of circulation depends on the droplet size and contact angle [39, 40]. Buoyancy driven internal circulation also may emerge during evaporation [41], but it is only relevant on superhydrophobic substrates and large droplets (Bond number away from the neighbourhood of 1). The internal advection is also noted to enhance the evaporation via interfacial shear induced augmentation of the surrounding Stefan flow [31, 38]. Particle Image Velocimetry (PIV) experiments have been conducted to understand the nature of the internal advection in droplets on inclined substrates, and the velocity contours and vector fields have been illustrated in fig. 6. The evaporative flux difference (between rear and front edge of TL) on inclined surfaces leads to morphing of the internal advection kinetics.

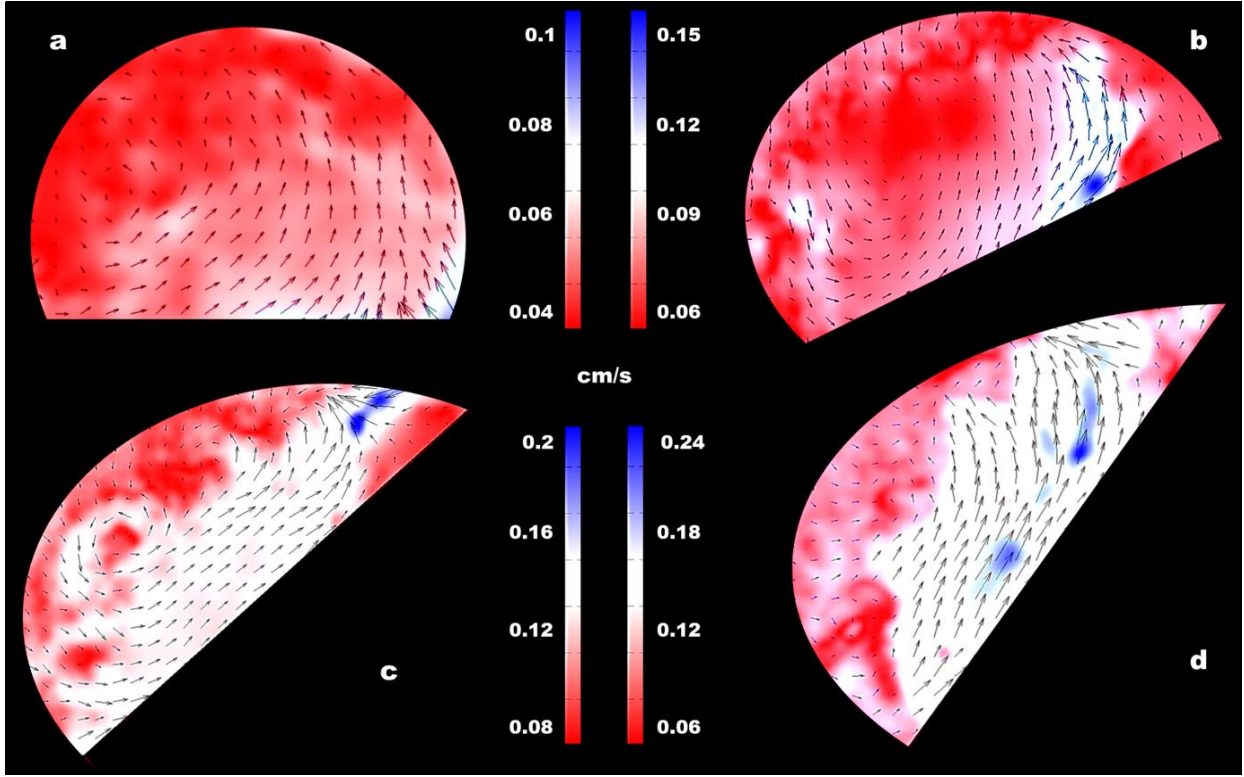

**Figure 6:** Velocity contours and vector field at the vertical mid-plane of the droplet on Teflon, for φ= (a) 0⁰ (b) 30⁰ (c) 45⁰ and (d) 60⁰.

On the horizontal surface, weak but consistent advection, spanning throughout the interior of the droplet is noted. As the declination increases, the advection velocity is observed to increase in magnitude, while simultaneously being confined more near the region of the droplet neighbouring the rear TL. At the same time, the strength of the circulation reduces in the front part of the droplet (towards the declination). It is noted that the formation of vortical flows are also pronounced due to declination (fig. 6 c). Evaporation of a droplet leads to evaporative cooling, which leads to thermal gradients across the droplet, generating thermal Marangoni advection [31, 38], which is noted within the droplet on the horizontal substrate. On an inclined substrate, the large value of the evaporative flux at the rear TL compared to the front TL leads to further asymmetric and non-uniform evaporative cooling, which augments the thermal gradient across the droplet. The relatively colder region near the rear TL compared to the front TL leads to the strengthening of the internal thermal Marangoni advection. Due to the nature of the asymmetric thermal gradient, the relative strength of the advection is stronger near the rear TL.

This behaviour is turn shears the droplet-air interface at the rear TL more compared to the front TL, which in turn aggravates the Stefan flow around the droplet in an asymmetric pattern. This replenishes the vapour diffusion layer shrouding the droplet surface with the ambient air, leading to further enhanced evaporation [31, 42-45]. To further the understanding, infrared thermography of the evaporating droplets is performed and is illustrated in Fig. 7. It is noted that the core of the droplet cools down to greater extents due to surface declination, which is an indication of the enhanced evaporation rates. For a sessile droplet with acute contact angle, the evaporative flux is symmetric about the vertical central axis. However, the flux is not uniform. Its strength is highest at the TL, and lowest at the apex of the droplet. This leads to non-uniform evaporative cooling (fig. 7, where the droplets show a central cooler region), leading to thermal gradients, which induces the thermal Marangoni circulation. The shape asymmetry on inclined surfaces distorts the symmetry of the evaporative flux over and above the already existent non-uniform nature (fig. 5). Also, the evaporation rate enhances. This is evident from the enhanced reduction in the core temperature and the size of the cooled core in case of droplets on inclined places (fig. 7, Teflon). This leads to strengthening of the thermal gradients across the droplet, which leads to enhanced thermal Marangoni circulation, which is evidenced from fig. 6.

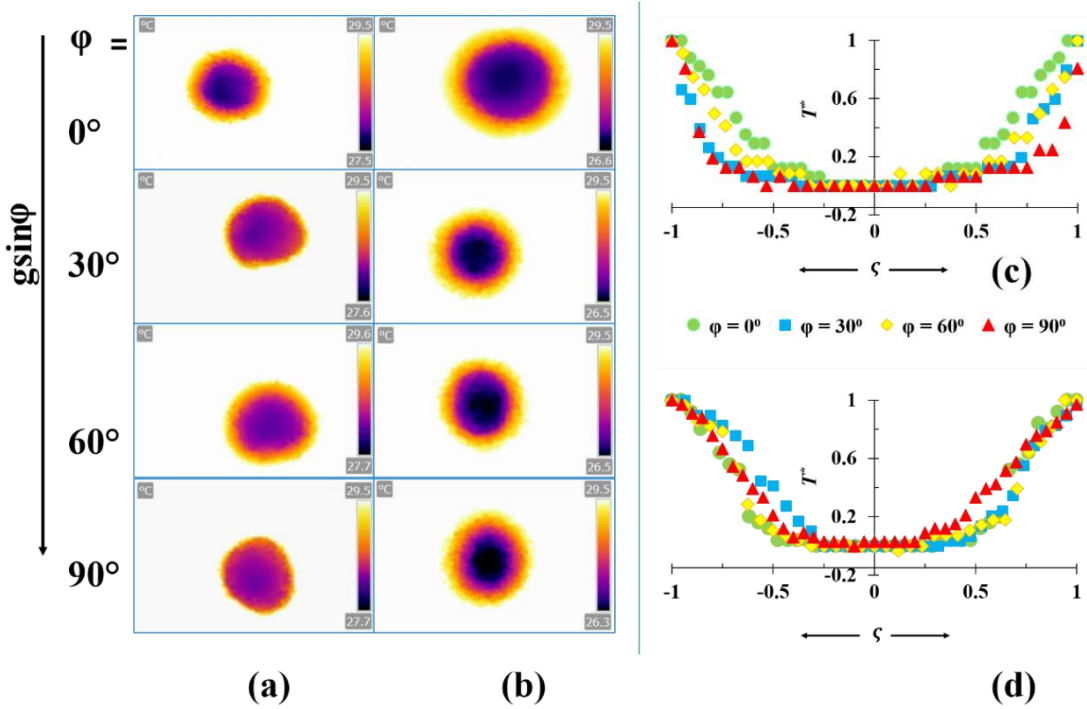

**Figure 7:** Infrared thermography illustrating temperature distribution within the droplets resting on (a) glass and (b) Teflon substrates at different declinations, (c) and (d) illustrate the non-dimensional temperature ($T^*$) map across the evaporating droplet on different inclined glass and Teflon, respectively. The positive ς axis represents the side containing the rear TL. The arrow (labelled $gsin\theta$ represents the direction of declination).

Fig. 7 (c) and (d) illustrate the internal non-dimensional temperature $T^* = \left( \dfrac{T - T_{\varsigma=0}}{T_{\varsigma=1} - T_{\varsigma=0}} \right)$ maps with respect to ς. It is noted that the side of the droplet with the rear TL exhibits steeper thermal gradients compared to the side containing the front TL. This shows that there is stronger advection induced mixing within the rear portion of the droplet compared to the front portion, which is in agreement with the observations from the flow visualization. To aid the understanding, a phase plot of the governing thermal Marangoni number (Ma) and the Rayleigh number (Ra) has been illustrated in fig. 8a. The complete methodology and mathematical formalism to determine the Ma and Ra and generate the phase plot with its stability regimes (marked 1 and 2) has been reported by the present authors in literature [31, 42]. In the plot, the region below the line 2 indicates weak thermal Marangoni circulation, the region in between 1 and 2 represents intermittently stable circulation, and the region above 1 represents unconditionally stable circulation. It is seen from the figure that with increasing declination, the circulation regime transits from the weak to intermittently stable, which is in agreement with the PIV observations. The Ma values are largely nearer to the critical Ma value (~81) for thermal Marangoni advection, than the Ra values are from the critical Ra (~1708) for thermal Rayleigh convection [31, 38]. Hence, the observed morphed internal advection is caused by the thermal Marangoni effect brought about by the change in evaporative cooling of the asymmetric shaped droplets. The spatio-temporally averaged circulation velocity is theoretically determined [31, 38, 42-45] from the thermal Marangoni advection model, and have been compared against the experimental velocities (fig. 8b), and good agreement between the two sets has been obtained.

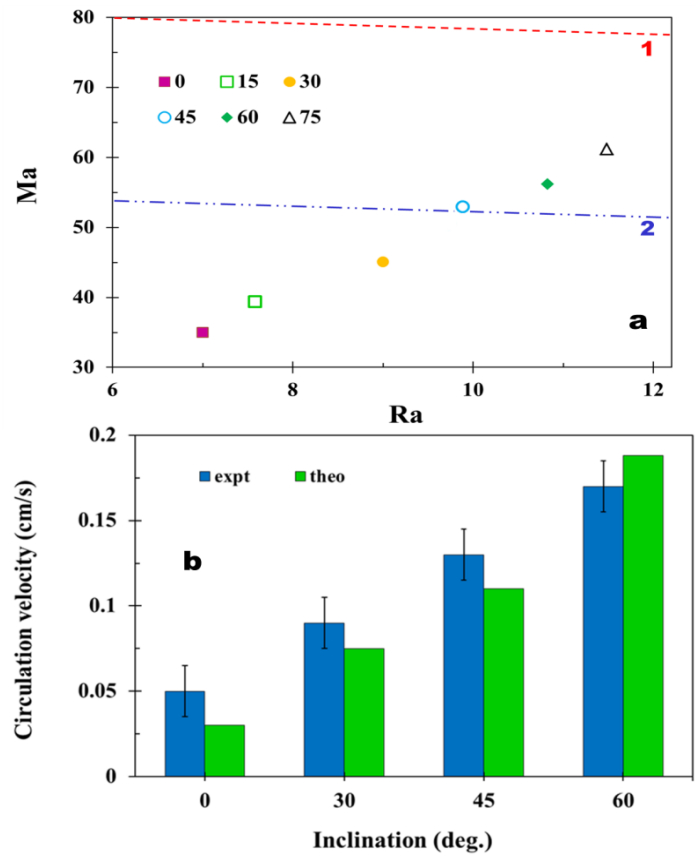

**Figure 8:** (a) Plot of the Ma vs the Ra for droplets evaporating on aluminium of different declination. The lines 1 and 2 represent the criteria for stable internal advection as proposed by Nield and Davis, respectively [31, 42]. (b) Comparison of the theoretical and experimental internal advection velocities.

**3. d. Pinning-depinning kinetics–role of declination and temperature**

The interplay of surface declination and temeprature has been studied for the aluminium substrates. Temperatures of 20 and 40 °C above the ambient temperature have been explored, and the volume evolution have been illustrated in fig. 9. It is noted that the interplay of declination and elevated temperatures only lead to appreciable acceleration of the evaporation for the ΔT=40 °C case. This can be explained based on the geometry of the tear shaped droplets. At zero declination, the change in evaporation is only due to the thermal stimulus. In case of φ=30°, the droplet shapes are still similar, and the evaporation rates are similar. However, at ΔT=40 °C, the droplet shapes play a major role. As the declination inrcreases, the rear TL thins out and leads of region of lower contact angle. The thermal stimulus is potent to drive rapid thin film evaporation in this region, which leads to enhanced evaporation rates at higher declination angles. The behavior of the non-dimensional contact diameter of the droplets on heated surfaces of different declinations has been illustrated in fig. 10. It is noted that the jump-stick phenomena of the droplets on aluminium is aggravted by the thermal stimulus, which further cements the previous discussion that the rear TL undergoes rapid thin film evaporation on inclined droplets. The typical values of the percentage change in wetting diameter during the thermal stimulus mediated jump-stick behavior have been tabulated in table S3.

Further, a normalised transition time ($τ_{tr}$) has been proposed to quantify the depinning dynamics of the droplets on inclined surfaces (fig. S5, supporting information). The $τ_{tr}$ is defined as the ratio of the time frame at which the droplet depinns for the first time from the CCR mode to the life time of the droplet. A droplet on horizontal aluminium shows the highesr $τ_{tr}$ (~0.88) compared to glass and Teflon (~0.53 and ~0.57, respectively). With increasing declination, the $τ_{tr}$ curves collapse (fig. S5 a) and become nearly independent of wettability, which illustrates the dominant role of the declination angle. As ΔT increases, the slope of the curve decreases (fig. S5 b) and at ΔT = 40ºC, the variation of $τ_{tr}$ with declination angle is weak. The near independency of $τ_{tr}$ with respect to higher temperatures indicates that the pinning force increases with increase in the temperature. Similar observations are also reported by on horizontal substrates [46, 47]. Increasing of temperature leads to reduction in the surface tension, and thus the same droplet which is at the verge of depinning on non-heated surface (eqn. 2) is now pinned due to decrease in the critical retention force.

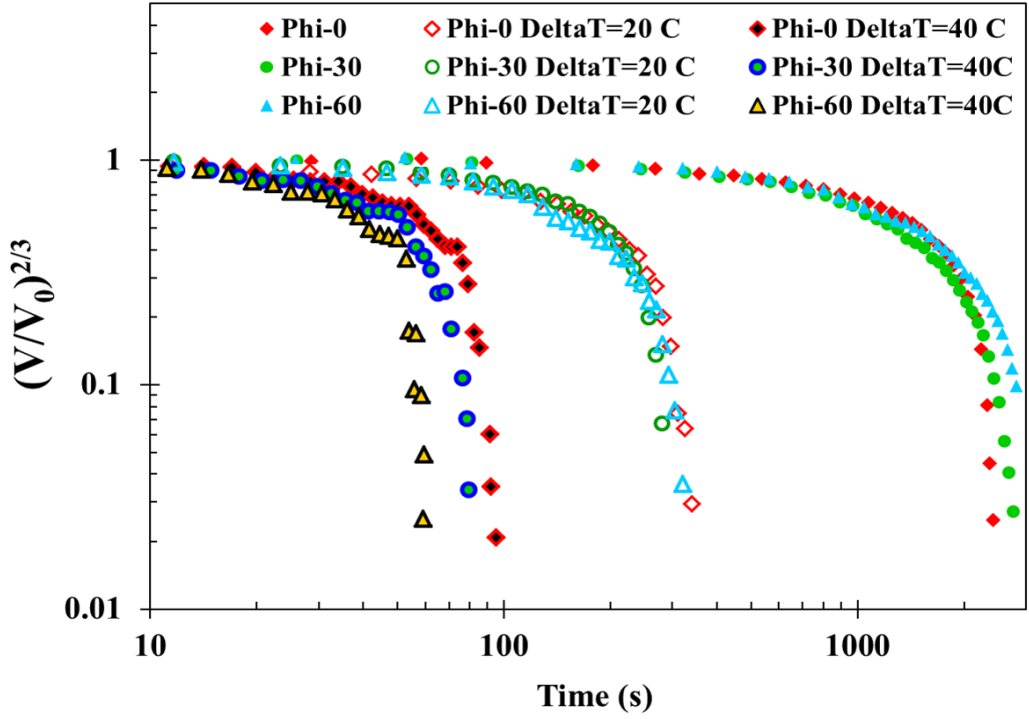

**Figure 9:** Non-dimensional volume vs time for droplets on aluminium substrates at different declinations and surface temperatures (ΔT=20 and 40°C).

A model to deduce the stick time for spherical cap droplet has been reported [48]. The stick time is defined as the time duration up to which the droplet evaporates in the CCR mode before the first jump appears. In the present study, the CCR mode appears more than once in inclined droplets, and hence, the time duration of the first CCR has been considered. The volume of a droplet from spherical cap geometry is

$$V = \pi R^3 \frac{\cos^3\theta - 3\cos\theta + 2}{3\sin^3\theta} \tag{6}$$

Where, R is the wetting radius and θ is the instantaneous average of the front and rear contact angles. The rate of the volume loss is as

$$\frac{dV}{dt} = \frac{\pi R^3}{(1+\cos\theta)^2}\left(\frac{d\theta}{dt}\right) + \pi R^2 \frac{\cos^3\theta - 3\cos\theta + 2}{\sin^3\theta}\left(\frac{dR}{dt}\right) \tag{7}$$

For pinning of the droplet dR/dt=0, and hence after integrating the remaining terms with respect to time, the stick time $t_{stick}$ for CCR mode is

$$t_{stick} = \frac{\pi R^3 (\Delta\theta)}{(1+\cos\theta_0)^2 (dV/dt)} \tag{8}$$

Where, $\theta_0$ is considered as the initial average of the front and rear contact angles for the present study and $R$ is the pinning radius of the droplet. $\Delta\theta$ is the difference between the $\theta_0$ and the average contact angle (say $\theta_t$) at which the TL depinning starts. $R, dV/dt, \theta_0$ and $\Delta\theta$ are determined from the experimental observations. The theoretical predictions are presented in table S3 and good match is noted in most cases.

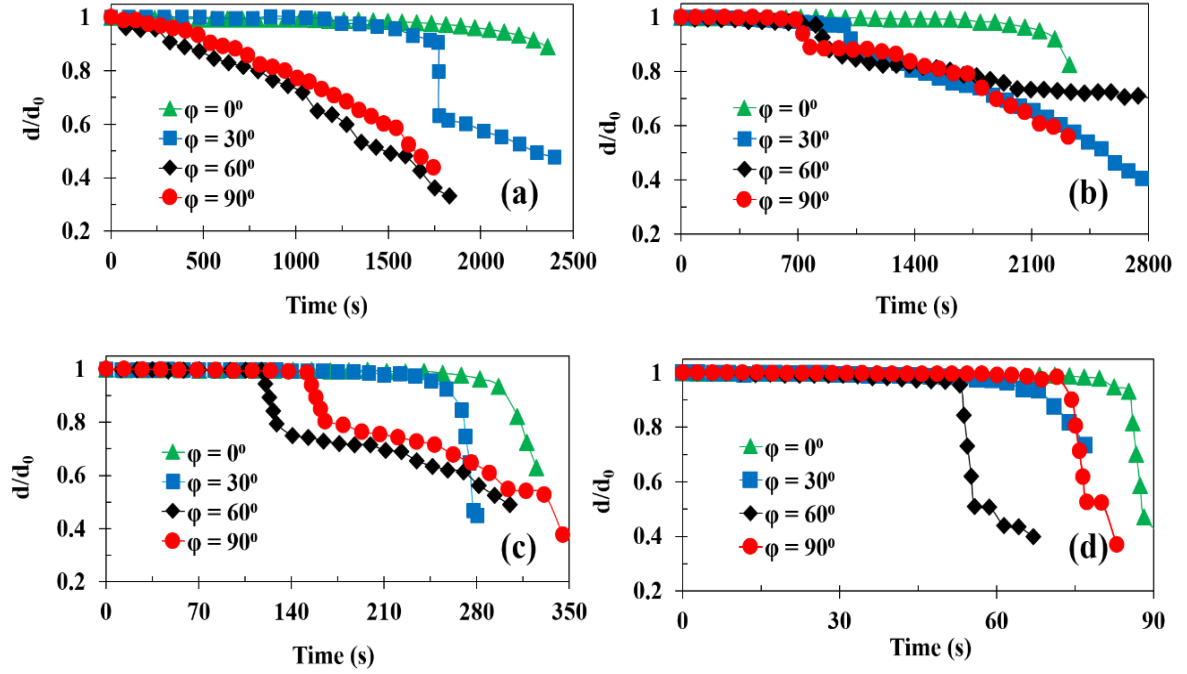

**Figure 10:** Dynamics of the contact diameter on (a) glass (b) aluminium (c) aluminium at $\Delta T = 20\ ^oC$ and (d) aluminium at $\Delta T = 40\ ^oC$.

### 3. e. Life time of asymmetric droplets

As the droplet life time is an important factor from utilitarian perspective, it has been evaluated for substrates at different declination. The nature of variation of droplet life times have been illustrated in Fig.10. For non-metallic substrates (glass and Teflon), the droplet life time varies inversely with respect to the initial contact angle dissimilarity. This can be explained as follows: at 30°, $\theta_f$ ~88.85° and $\theta_r$ ~82.30° are larger than the contact angle at φ=0° (θ ~73.57°). This causes decrease of the evaporative flux at both the edges of the TL, and at φ=30°, earlier depinning of the TL enhances this decrement further resulting in longer life time at φ=0°. For the heated substrate cases, the variation of the life time can be explained as the interplay between the contact angle dissimilarity and the onset of depinning of the TL. At $\Delta T$=20°C and φ=0°, θ ~87.31° (Table S2) and at φ=30°, $\theta_f$ ~74.84° and $\theta_r$ ~69.98°, which are lesser than the horizontal case, resulting in increased evaporation near both the edges of the TL, which ultimately reduces the life time. We note the interplay of wetting and shape asymmetry as: if the contact angle is small, the evaporation rate is high, whereas earlier depinning of the droplet reduces the evaporative flux. This is further distinct when the jump in the TL occurs, which suddenly decreases the wetting diameter and increases the contact angles. On inclined surfaces, these two regimes compete, and the effective droplet life time thus shows non-linear trends.

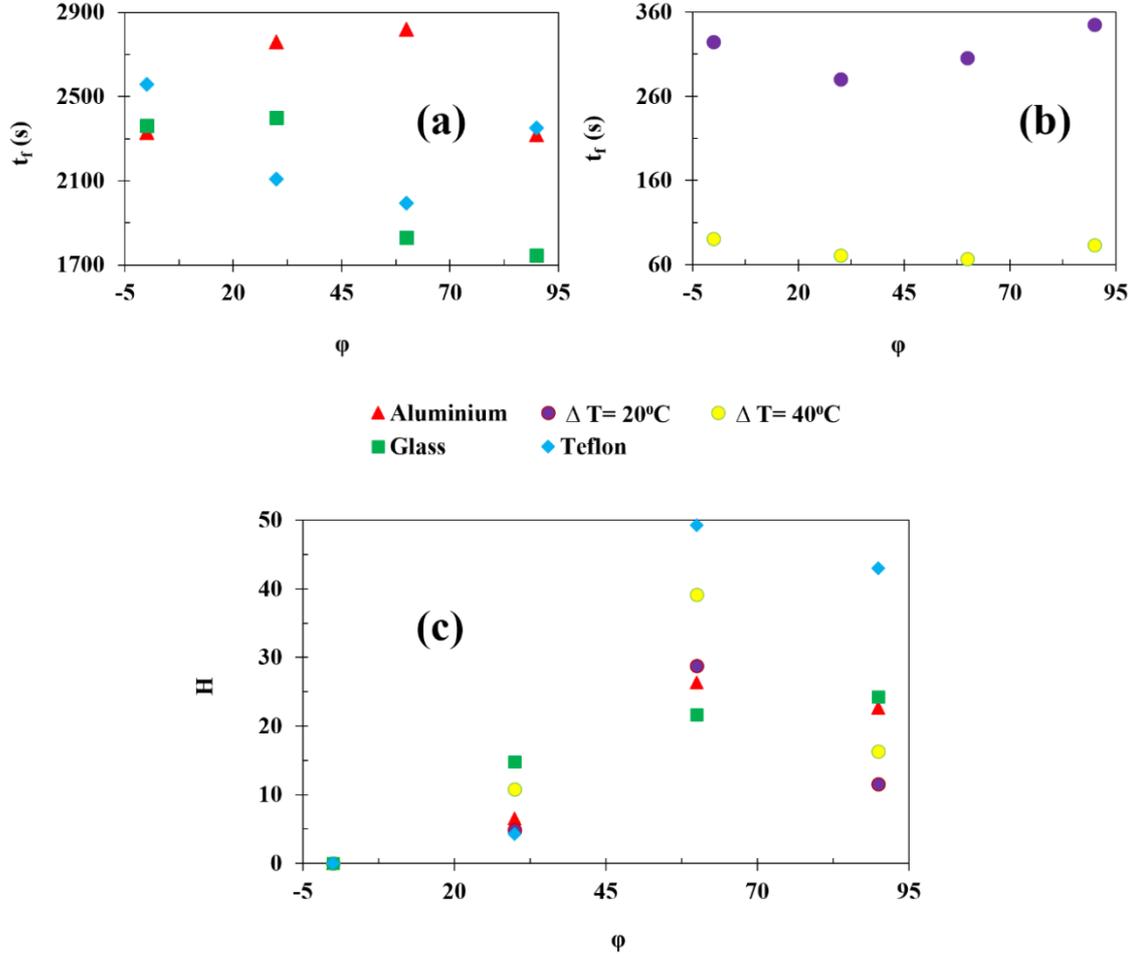

**Figure 10:** Variation of the droplet lifetime with declination angle on various substrates for (a) non–heating (b) heating case (only aluminium substrate) and (c) variation of contact angle dissimilarity with the substrate declination angles.

The droplet evaporates in the CCR, CCA, mixed or stick-jump modes, or a combination of these. For a generalised case, the droplet life time (t) comprises the time duration of all four modes and is expressed as

$$t = t_{CCR} + t_{CCA} + t_{mixed} + t_{jump} \tag{9}$$

The droplet evaporation time in the CCR mode is expressed as [49]

$$t_{CCR} = \frac{\rho R_0^2}{D(C_s - C_\infty)} \int_{\theta^*}^{\theta_0} \frac{d\theta}{g(\theta)} \tag{10}$$

Where, $\rho$ is the density of the fluid. $R_0$ is the initial base radius of the droplet, $\theta^*$ is the average of the $\theta_f$ and $\theta_r$ at the moment when the first CCR mode terminates and depinning of the TL starts. The semi-analytical solution technique is reported [49] to solve the integral in equation (10), and $g(\theta)$ can be expressed as

$$g(\theta) = \frac{(\pi - \theta)^3}{0.1139\theta^4 - 1.0009\theta^3 + 3.4838\theta^2 - 6.0419\theta + 6.0233} \quad (11)$$

Where, θ is in radians. The droplet evaporation time in CCA mode can be determined as [27]

$$t_{CCA} = \frac{\rho R_0^2}{D(C_s - C_\infty)} \frac{(2 + \cos\theta^*)\sin\theta^*}{2g(\theta^*)} \quad (12)$$

In mixed mode, the droplet will evaporate with decrease in contact angles as well as wetting diameter simultaneously. It is assumed that the evaporation time in mixed mode has the equal effect of both CCR and CCA modes. So $t_{mixed}$ is be evaluated by averaging $t_{CCR}$ and $t_{CCA}$ pertinent to the mixed regime of evaporation process. The evaporation time in stick-jump mode has been neglected as it is an instantaneous process and its duration is negligible compared to the other modes. From eqn. 9, the life time t(π/2) of a hemispherical drop (contact angle θ=π/2) is as

$$t\left(\frac{\pi}{2}\right) = \frac{\rho(3V_0\pi^{1/2})^{2/3}}{10D(C_s - C_\infty)} \quad (13)$$

The above relation has been considered as the reference to non-dimensionalize the life time (τ) of the droplets in all cases as

$$\tau = t \Bigg/ \left\{ \frac{\rho(3V_0\pi^{1/2})^{2/3}}{10D(C_s - C_\infty)} \right\} \quad (14)$$

The theoretical values from eqn. 14 are illustrated in Fig.11. The experimental values are non-dimensionalized using the values from literature [49]. The theoretically predicted values agree well with the experimentally noted values.

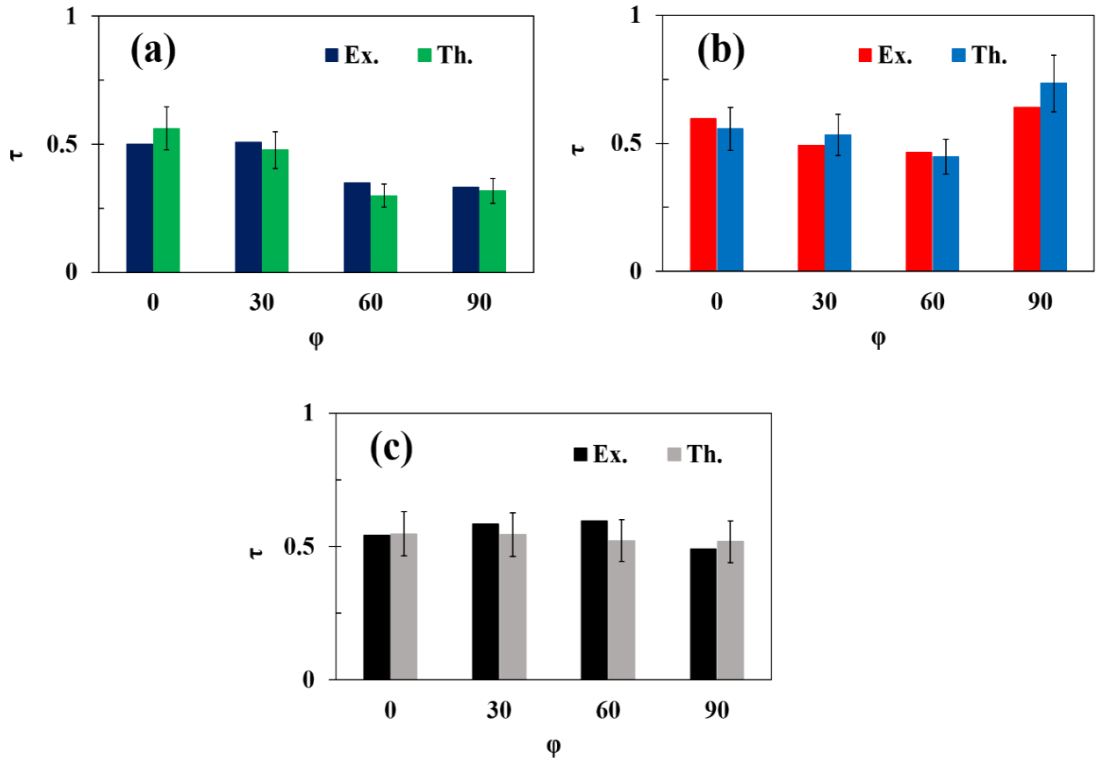

**Figure 11:** The nature of variation of normalized life time of the droplet (ex. and th. Represent experimental and theoretical, respectively) vs. the declination angles for (a) glass (b) Teflon and (c) aluminium substrates.

## 4. Conclusions

The major conclusions from the present study can be summarized as follows:

1. Sessile droplets with substrate declination exhibit distorted shape and evaporate at different rates compared to droplets on the same horizontal substrate. The front and rear contact angles appear, and evolve differently compared to the horizontal case. The changes in regimes of evaporation appear more often on inclined surfaces. In several cases, the slip-stick and jump-stick modes are prominent during evaporation.
2. Shape of sessile droplets causes non-uniformity (but symmetric) in the evaporative flux. For droplets on inclined substrates, the flux is also asymmetric. Due to smaller contact angle at the rear TL, it is the zone of a higher evaporative flux. The minima of the flux is also shifted away from the centre of the droplet.
3. The velocimetry shows the increased internal circulation velocity within the inclined droplets. The circulation is more prominent near the read of the droplet compared to the frontal part. Asymmetry in the evaporative flux leads to higher temperature gradients, which ultimately enhances the thermal Marangoni circulation near the rear of the droplet where the evaporative flux is highest.
4. A stability map for the thermal Marangoni number and the Rayleigh number shows that the internal circulation is dominantly morphed by the changes in the thermal Marangoni effect. A model adopted from previous reports by the authors is used to predict the thermal Marangoni advection velocity, and good match is obtained.

5. The declination angle and imposed thermal conditions interplay and lead to morphed evaporation kinetics than droplets on horizontal heated surfaces. This is caused by the rapid film evaporation from the rear end of the tear shaped droplet. Pinning and depinning kinetics of the TL is noted to govern the evaporation from heated inclined surfaces.
6. Even weak movements of the TL alter the evaporation dynamics significantly, by changing the shape of the droplet from ideally elliptical to almost spherical cap, which ultimately reduces the evaporative flux.
7. The droplet life time is noted to be inversely proportional to the initial contact angle dissimilarity. The life time of the droplet is modelled by modifying available models for non-heated substrate, to account for the shape asymmetry. Good match with the experimental values is obtained.

The present findings may find strong implications towards design and development of systems involving microscale thermo-hydrodynamics.

## Acknowledgements

PD thanks IIT Kharagpur for partially funding the research work (vide ISIRD project; code SFI).

## Supporting information document

This document contains additional details on the surface and fluid interaction parameters, illustrations on the slip-stick dynamics, evaporation life times, parameter tables, etc.